\documentclass[11pt,epsf,letterpaper]{article}%
\usepackage{setspace}
\usepackage{amsmath}
\usepackage{amsfonts}
\usepackage{verbatim}
\usepackage{amssymb}
\usepackage{graphicx}%
\providecommand{\U}[1]{\protect\rule{.1in}{.1in}}
\begin{document}

\title{Classification of Six Derivative Lagrangians of Gravity and Static
Spherically Symmetric Solutions}
\author{Julio Oliva$^{1}$ and Sourya Ray$^{2}$\\$^{1}$Instituto de F\'{\i}sica, Facultad de Ciencias, Universidad Austral de
Chile, Valdivia, Chile.\\$^{2}$Centro de Estudios Cient\'{\i}ficos (CECS), Casilla 1469, Valdivia, Chile.\\{\small {CECS-PHY-10/04}}\\{\small {julio.oliva@docentes.uach.cl, ray@cecs.cl}}}
\maketitle

\begin{abstract}
We classify all the six derivative Lagrangians of gravity, whose traced field
equations are of second or third order, in arbitrary dimensions. In the former
case, the Lagrangian in dimensions greater than six, reduces to an arbitrary
linear combination of the six dimensional Euler density and the two linearly
independent cubic Weyl invariants. In five dimensions, besides the independent
cubic Weyl invariant, we obtain an interesting cubic combination, whose field
equations for static spherically symmetric spacetimes are of second order. In
the later case, in arbitrary dimensions we obtain two combinations, which in
dimension three, are equivalent to the complete contraction of two Cotton
tensors. Moreover, we also recover all the conformal anomalies in six
dimensions. Finally, we present the general static, spherically symmetric
solution for some of these Lagrangians.

\end{abstract}
\tableofcontents

\bigskip

\section{Introduction}

Einstein's General Relativity is not only the most successful classical theory
of gravity in four dimensions, it is also the simplest theory possessing some
nice properties. The two of the most important characteristics being general
covariance and second order field equations. In fact, it was shown by Lovelock
that, in four dimensions, General Relativity is the unique generally covariant
theory of gravity (upto an addition of a cosmological constant) which gives
second order equations of motion \cite{Lovelock}. However, in higher
dimensions, there exists higher curvature theories, namely the Lovelock
theories, which also gives second order field equations. These theories are
generically characterized by higher curvature invariants in the action. At
each order $k$, the combination of higher curvature invariants is unique, the
integral of which on a compact manifold of dimension $2k$ gives the Euler
characteristic of the manifold. There are also other interesting
characteristics of the Lovelock class of theories. Primarily, exact analytic
black hole solutions are known to exist \cite{BD,JTW} (see also
\cite{GG,CHARM} and references therein). Black holes are widely believed to
exist in nature as a final state of a gravitational collapse.\ They are also
the simplest objects to study in any gravitational theory. Therefore, exact
black hole solutions are of significant importance. More recently, in the
context of AdS/CFT correspondence, exact asymptotically AdS black hole
solutions in a gravity theory have been proven to be useful in studying the
holographic properties of a finite temperature conformal field theory on the
boundary \cite{WITT}. Secondly, the Lovelock theories also admit Birkhoff's
theorem, which states that any solution which has spherical, planar or
hyperbolic symmetry must be locally isometric to the corresponding static
black hole solution \cite{ZEG}. Generically, the admittance of Birkhoff's
theorem suggests the lack of spin-$0$ mode excitations in the linearized field equations.

There are also other theories of gravity which admit exact analytic black hole
solutions and further admit Birkhoff's theorem. These theories, being outside
the Lovelock class, are generically higher derivative theories and
consequently possess ghost degrees of freedom. One well known higher
derivative theory is the conformal theory of gravity in four dimensions, which
is obtained from an action quadratic in the conformal Weyl tensor. The action
is thus invariant under Weyl rescalings and the field equations are traceless.
Birkhoff's theorem in conformal gravity states that modulo a conformal factor
the most general spherically symmetric solution is static \cite{Riegert}.
Considering the same action in dimensions other than four, one loses the
property of invariance under Weyl rescalings. Even then, the theory admits
exact analytic black hole solutions with spherical, planar or hyperbolic
symmetry, and further admits Birkhoff's theorem \cite{birk}. Note that, in
arbitrary dimensions, though the field equations are of fourth order, the
trace of the field equations are of order two. This can easily be seen as
follows. Varying the action gives
\begin{equation}
\delta I:=\int\delta(\sqrt{-g}\mathcal{L})=\int\sqrt{-g}\mathcal{E}_{\mu\nu
}\delta g^{\mu\nu}%
\end{equation}
Now, consider infinitesimal Weyl-rescalings of the metric $\delta g^{\mu\nu
}=\lambda g^{\mu\nu}$. Under such variations, the Lagrangian will vary as
$\lambda(2-D/2)\sqrt{-g}\mathcal{L}$, which implies $\mathcal{E}_{\ \mu}^{\mu
}=({\frac{4-D}{2}})\mathcal{L}$. This shows that the trace of the field
equations, being proportional to the Lagrangian density, must be of second order.

Recently, a theory of massive gravity was constructed in three dimensions
\cite{bht}, where the Lagrangian is a particular combination of quadratic
curvature invariants given by
\begin{equation}
K=R_{ab}R^{ab}-\frac{3}{8}R^{2}. \label{KNMG}%
\end{equation}
Again, the theory admits exact analytic black hole solutions \cite{OTT}. The
Lagrangian of this theory has a unique property in three dimensions, that the
field equations have a second order trace. So, it is natural to wonder if
there could be other theories of gravity which, although non-realistic, have
some special properties which allows one to obtain exact analytic black hole
solutions and can serve as toy models of gravitational theories. Specifically,
it might be useful to classify higher-derivative theories of gravity whose
traced field equations have a reduced order.

In this paper we construct the most general Lagrangian which is a linear
combination of scalars of the form
\begin{equation}
R_{....}R_{....}R_{....}\text{ and }\nabla_{.}R_{....}\nabla_{.}R_{....}\ ,
\label{terms}%
\end{equation}
which are characterized by the number of derivatives of the metric (hereafter,
the degree of differentiation) $n=6$, such that the trace of the field
equations is of order three or less.

We will show that when the trace is resticted to be of order two, then, in
dimensions six or higher, there are only three linearly independent possible
invariants which have a second order trace. They are the six-dimensional Euler
density, and the two linearly independent scalars constructed by contracting
all the indices of three conformal Weyl tensors. However, in five dimensions,
we obtain a peculiar independent invariant which can be thought of as a cubic
generalization of (\ref{KNMG}) and has been studied separately in \cite{OR1}
and \cite{QTG,QTGHOL}. We also obtain the general static spherically symmetric
solution for some of these theories. Based on our analysis for six-derivative
theories, we present a conjecture classifying all the scalars of arbitrary
order, which give second order traced field equations in various dimensions.

When the trace is restricted to be of order three, in arbitrary dimensions, we
obtain two additional scalars. These two scalars are not independent in
dimensions three and six. In six dimensions they reduce to one of the
conformal anomalies.

One future direction of study is to see which of these theories admit a
Birkhoff's theorem.

In Section \ref{quadratic}, for completeness we review the $n=4$ case. The
case $n=6$ is analyzed in Section \ref{cubic}. In Section
\ref{Exact Solutions} we focus our attention on obtaining the general static,
spherically symmetric solution for some of the theories defined in Section
\ref{cubic} in arbitrary dimensions.

\section{Quadratic Combinations. $n=4$}

\label{quadratic}

In this section, we review how to obtain the most general quadratic
Lagrangian, having second order traced field equations \cite{farhoudi,naoda}.

The most general quadratic combination of curvature invariants in arbitrary
dimension is given by\footnote{Note that the only non-quadratic term with
degree of differentiation $4$ is $\square R$, which is boundary term.}
\begin{equation}
\mathcal{L}_{Q}:=a\ R^{abcd}R_{abcd}+b\ R^{ab}R_{ab}+c\ R^{2}\ , \label{LQ}%
\end{equation}
where $a,b$ and $c$ are arbitrary constants. The trace of the field equations
coming from this Lagrangian are%
\begin{equation}
G_{a}^{(2)a}=\left(  4a+\frac{D}{2}b+2\left(  D-1\right)  c\right)  \nabla
_{a}\nabla^{a}R-4a\nabla_{a}\nabla_{b}R^{ab}-\frac{D-4}{2}\mathcal{L}_{Q}.
\label{trQua}%
\end{equation}
Imposing the trace $G_{a}^{(2)a}$ to be of second order, implies that the
coefficients $a,b$ and $c$ must be chosen such that the first two terms in
(\ref{trQua}) vanish, i.e.,%
\begin{equation}
\nabla_{a}\nabla_{b}\left[  -4aR^{ab}+g^{ab}\left(  4a+\frac{D}{2}%
b+2(D-1)c\right)  R\right]  =0. \label{seedqua}%
\end{equation}
The above equation is satisfied only if the term inside the bracket is
proportional to the Einstein tensor, which is the most general divergenceless,
symmetric, rank two tensor linear in the curvature\footnote{Since a
divergenceless vector $J^{a}$ cannot be constructed locally out of the
curvature, the equation $\nabla_{a}J^{a}=0$, with $J^{a}:=\nabla_{b}\left[
-4aR^{ab}+g^{ab}\left(  4a+\frac{D}{2}b+2(D-1)c\right)  R\right]  $, does not
have any non-trivial solution.}. Consequently, the coefficients in
(\ref{seedqua}) must fulfill%
\begin{equation}
-4a=\gamma\ \ \ \ \text{and}\ \ \ \ 4a+\frac{D}{2}b+2(D-1)c=-\frac{\gamma}%
{2}\ . \label{constraintsquadratic}%
\end{equation}
Since there are four variables ($a,b,c,\gamma$) and two equations, there is a
bi-parametric family of solutions given by%
\begin{equation}
\mathcal{L}_{Q}=-\frac{\gamma}{4}R^{abcd}R_{abcd}+b\ R^{ab}R_{ab}+\frac
{\gamma-bD}{4(D-1)}R^{2}\ . \label{LQ2}%
\end{equation}
One can further factor out the four dimensional Euler density to write
(\ref{LQ2}) in the form%
\begin{equation}
\mathcal{L}_{Q}=\alpha\mathcal{N}_{4}+\beta\mathcal{E}_{4}\ ,
\label{quadratic1}%
\end{equation}
where $\alpha=b-\gamma$ and $\beta=-\frac{\gamma}{4}$ are arbitrary constants,
$\mathcal{N}_{4}$ is defined by%
\begin{equation}
\mathcal{N}_{4}:=4R^{ab}R_{ab}-\frac{D}{(D-1)}R^{2}={\frac{1}{2^{4}}}\left(
\frac{D-2}{D-3}\right)  \delta_{c_{1}d_{1}c_{2}d_{2}}^{a_{1}b_{1}a_{2}b_{2}%
}\left(  C_{a_{1}b_{1}}^{\ \ \ c_{1}d_{1}}C_{a_{2}b_{2}}^{\ \ \ c_{2}d_{2}%
}-R_{a_{1}b_{1}}^{\ \ \ c_{1}d_{1}}R_{a_{2}b_{2}}^{\ \ \ c_{2}d_{2}}\right)
\ , \label{F4}%
\end{equation}
and $\mathcal{E}_{4}:=R^{2}-4R_{ab}R^{ab}+R_{abcd}R^{abcd}$ is the
Gauss-Bonnet combination which corresponds to the four dimensional Euler
density. Thus, we have shown that for combinations quadratic in the curvature,
the most general Lagrangian which has second order traced field equation, can
be expressed as a linear combination of the four dimensional Euler density
$\mathcal{E}_{4}$ and $\mathcal{N}_{4}$ defined in Eq. (\ref{F4}).

In dimensions higher than three, one can further use the following relation%
\begin{equation}
\label{keyrelation}C^{abcd}C_{abcd}=\mathcal{E}_{4}+\left(  \frac{D-3}%
{D-2}\right)  \mathcal{N}_{4}\ ,
\end{equation}
where $C_{abcd}$ is the Weyl tensor and $C^{abcd}C_{abcd}$ is the
four-dimensional conformal Weyl invariant. This means that for dimensions
higher than three, $\mathcal{L}_{Q}$ in (\ref{LQ2}) can be equivalently
expressed as a linear combination of the four-dimensional Weyl invariant and
the Euler density.

\section{Lagrangians with $n=6$}

\label{cubic}

In this section, we generalize the previous discussion for Lagrangians of
degree of differentiation six in arbitrary dimensions $D$. We start by
considering a generic combination of the twelve linearly independent
\cite{fulling}, curvature invariants of degree six
\begin{equation}
\mathcal{L}=\sum_{i=1}^{12}A^{i}L_{i}\ , \label{Lcubic}%
\end{equation}
where $A^{i}$'s are arbitrary coefficients and $L_{i}$'s are given by%
\begin{align}
&  L_{1}=R^{abcd}R_{cdef}R_{\ \ ab}^{ef},\,L_{2}=R_{\ \ cd}^{ab}%
R_{\ \ bf}^{ce}R_{\ \ ae}^{df},\,L_{3}=R^{abcd}R_{cdbe}R_{\ a}^{e}%
,\,L_{4}=RR^{abcd}R_{abcd},\,L_{5}=R^{abcd}R_{ac}R_{bd},\,\nonumber\\
&  L_{6}=R^{ab}R_{bc}R_{\ a}^{c},\,L_{7}=RR^{ab}R_{ab},\,L_{8}=R^{3}%
,\,L_{9}=\nabla_{a}R\nabla^{a}R,\,L_{10}=\nabla_{a}R_{bc}\nabla^{a}%
R^{bc},\,L_{11}=\nabla_{p}R_{abcd}\nabla^{p}R^{abcd},\nonumber\\
&  L_{12}=\nabla_{a}R_{bc}\nabla^{b}R^{ac}. \label{Ls}%
\end{align}
Note however that when we neglect a total derviative the twelve terms are not
linearly independent, as one can write two of the invariants in terms of the
other ten in the following way\footnote{We would like to thank Nicolas
Boulanger for pointing this out to us.}%
\begin{align}
L_{11}  &  =2L_{1}-4L_{2}+2L_{3}-4L_{5}+4L_{6}-L_{9}+4L_{10}+\nabla_{a}%
\nabla_{c}[2R^{abde}R_{\ bde}^{c}-8R^{abcd}R_{bd}+8RR^{ac}\nonumber\\
&  -12R^{ab}R_{\ b}^{c}+2g^{ac}(2R^{bd}R_{bd}-R^{2})]\nonumber\\
L_{12}  &  =L_{5}-L_{6}+{\frac{1}{4}}L_{9}+\nabla_{a}\nabla_{c}[R_{\ b}%
^{a}R^{bc}-RR^{ac}+{\frac{1}{4}}g^{ac}R^{2}]
\end{align}
Using the above relations, one can rewrite the Lagrangian as a linear
combination of only ten curvature invariants with coefficients $\tilde{A}^{i}$
($i=1,\cdots,10$). Extremizing the action constructed with the Lagrangian
(\ref{Lcubic}) with respect to the metric gives the field equations%
\begin{equation}
\mathcal{G}_{ab}^{(3)}:=\sum_{i=1}^{10}\tilde{A}^{i}G_{\left(  i\right)
ab}^{(3)}=0\ ,
\end{equation}
where $G_{\left(  i\right)  ab}^{(3)}$ are defined respectively in equations
(\ref{Gab1}-\ref{Gab8}) in the Appendix \ref{eom}. Now, requiring the trace
$\mathcal{G}_{a}^{(3)a}$ to be of order three, is the same as imposing trace
to be proportional to $\mathcal{L}$ (See Appendix \ref{tofe}). This gives us a
set of eight equations (\ref{SystemD}) for the twelve variables ($A^{i}$) and
a parameter $u$ (analoguous to $\gamma$ in the quadratic case
(\ref{constraintsquadratic})). Solving these equations for arbitrary
dimensions, we obtain a five-parameter family of solutions. The details of the
equations and its solution are given in the Appendix \ref{eom}.
This implies that in $D>6$, there are five linearly independent curvature
invariants (of degree of differentiation six) which gives rise to third (or
lower) order traced field equations. They are as follows. Firstly, the
six-dimensional Euler density given by
\begin{equation}
\mathcal{E}_{6}:=2L_{1}+8L_{2}+24L_{3}+3L_{4}+24L_{5}+16L_{6}-12L_{7}+L_{8}\ ,
\end{equation}
obviously gives second-order traced field equations. Secondly, there are two
independent algebraic invariants constructed out of three Weyl tensors, namely
$W_{1}=C_{\ \ cd}^{ab}C_{\ \ ef}^{cd}C_{\ \ ab}^{ef}$ and $W_{2}%
=C_{abcd}C^{ebcf}C_{\ ef}^{a\ \ d}$, which also gives second order traced
filed equations. These two Weyl invariants are given in terms of the $L_{i}$'s
as%
\begin{align}
W_{1}  &  =L_{1}+\dfrac{12}{D-2}L_{3}+\dfrac{6}{(D-1)(D-2)}L_{4}+\dfrac
{24}{(D-2)^{2}}L_{5}\nonumber\\
&  +\dfrac{16(D-1)}{(D-2)^{3}}L_{6}-\dfrac{24(2D-3)}{(D-1)(D-2)^{3}}%
L_{7}+\dfrac{8(2D-3)}{(D-1)^{2}(D-2)^{3}}L_{8} \label{W1}%
\end{align}
and
\begin{align}
W_{2}  &  =-\dfrac{1}{4}L_{1}+L_{2}+\dfrac{3}{D-2}L_{3}+\dfrac{3}%
{2(D-1)(D-2)}L_{4}+\dfrac{3D}{(D-2)^{2}}L_{5}\nonumber\\
&  +\dfrac{2(3D-4)}{(D-2)^{3}}L_{6}-\dfrac{3(D^{2}+D-4)}{(D-1)(D-2)^{3}}%
L_{7}+\dfrac{(D^{2}+D-4)}{(D-1)^{2}(D-2)^{3}}L_{8}. \label{W2}%
\end{align}
In addition, there are two other curvature invariants $\Sigma$ and $\Theta$
listed in equations (\ref{Sigma}) and (\ref{Theta}) respectively which give
third order traced field equations. However, in dimensions $D\leq6$, the above
curvature invariants are not all linearly independent. For example, in $D=3$
and $6$, the invariants $\Sigma$ and $\Theta$ are proportional to each other
modulo a total derivative. It is interesting to note that, in six dimensions,
requiring the traced field equations to be of third order (or less), we
recover all the four ($1$ type-A and $3$ type-B) non-trivial conformal
anomalies \cite{Cites On Anomaly1}-\cite{Cites On Anomaly3}. In Table I below,
we list all the curvature invariants in dimensions greater than or equal to
$3$, which lead to third (or less) order traced field equations.

\begin{table}[ptbh]
\begin{center}%
\begin{tabular}
[c]{|c|c|c|c|c|c|}\hline
$G_{a}^{\ a}$ & $D=3$ & $D=4$ & $D=5$ & $D=6$ & $D>6$\\\hline
$\partial^{2}g$ & $\nexists$ & $W_{1}\sim W_{2}$ & $W_{1}\sim W_{2}$,
$\mathcal{N}_{6}$ & $W_{1},\ W_{2}$ & $W_{1},\ W_{2},$ $\mathcal{E}_{6}%
$\\\hline
$\partial^{3}g$ & $C_{abc}C^{abc}$ & $\Sigma,\ \Theta$ & $\Sigma,\ \Theta$ &
$\Sigma\sim\Theta$ & $\Sigma,\ \Theta$\\\hline
\end{tabular}
\end{center}
\caption{Here $C_{abc}$ denotes the Cotton tensor. Note that in dimension five
a new combination $\mathcal{N}_{6}:=-24L_{3}-\frac{21}{4}L_{4}-40L_{5}%
-\frac{320}{9}L_{6}+\frac{97}{3}L_{7}-\frac{31}{9}L_{8}$ appears.}%
\end{table}Returning to the set of invariants that gives second order traced
field equations we find that in dimensions $D\neq5$, they are spanned by the
basis set
$\{\mathcal{E}_{6},W_{1},W_{2}\}$ up to a total derivative. However, in $D=5$
this is not the case. In particular, there exists a \textquotedblleft special"
linearly independent invariant which generalizes $\mathcal{N}_{4}$ to the
cubic case. This is realized by noting that the following relation is
analogous to equation (\ref{keyrelation})
\begin{equation}
4W_{1}+8W_{2}=\mathcal{E}_{6}+\left(  \frac{D-5}{D-2}\right)  \mathcal{N}_{6}
\label{relforN6}%
\end{equation}
where
\begin{align}
\mathcal{N}_{6}:=  &  -24L_{3}-\frac{3\left(  D+2\right)  }{(D-1)}L_{4}%
-\frac{24D}{D-2}L_{5}-\frac{16D(D-1)}{(D-2)^{2}}L_{6}+\frac{12(D^{3}%
-2D^{2}+6D-8)}{(D-2)^{2}(D-1)}L_{7}\nonumber\label{defineN6}\\
&  -\frac{(D^{4}-3D^{3}+10D^{2}+4D-24)}{(D-2)^{2}(D-1)^{2}}L_{8}%
\end{align}
is the cubic counterpart of $\mathcal{N}_{4}$. Let us rewrite equation
(\ref{relforN6}) in the form
\begin{align}
\mathcal{N}_{6}  &  :={\frac{D-2}{D-5}}(4W_{1}+8W_{2}-\mathcal{E}_{6})\\
&  ={\frac{1}{2^{3}}}\left(  \frac{D-2}{D-5}\right)  \delta_{c_{1}d_{1}%
c_{2}d_{2}c_{3}d_{3}}^{a_{1}b_{1}a_{2}b_{2}a_{3}b_{3}}\left(  C^{\ \ \ c_{1}%
d_{1}}_{a_{1}b_{1}}C^{\ \ \ c_{2}d_{2}}_{a_{2}b_{2}}C^{\ \ \ c_{3}d_{3}%
}_{a_{3}b_{3}}-R^{\ \ \ c_{1}d_{1}}_{a_{1}b_{1}}R^{\ \ \ c_{2}d_{2}}%
_{a_{2}b_{2}}R^{\ \ \ c_{3}d_{3}}_{a_{3}b_{3}}\right)
\end{align}
The term inside the parenthesis on the right hand side vanishes identically in
dimensions lower than five, since for $D\leq5$
\begin{equation}
\mathcal{E}_{6}=4W_{1}+8W_{2}\equiv0.
\end{equation}
However, in $D=5$ (and greater than $5$) this gives a non-vanishing invariant
as can be seen by expressing $W_{1}$, $W_{2}$ and $\mathcal{E}_{6}$ in terms
of $\{L_{1},\cdots,L_{8}\}$, thereby obtaining the expression (\ref{defineN6}%
). This imples that in $D=5$, the basis is $\{W_{1}\left(  \sim W_{2}\right)
,\mathcal{N}_{6}\}$ up to a total derivative. Similar results have been found
for quartic invariants (see Appendix B of reference \cite{OR1}) i.e., in
dimensions $D\neq7$, any invariant giving second order traced field equations
can be expressed as a linear combination of the eight-dimensional Euler
density, and all the linearly independent Weyl invariants, however, in $D=7$
there is an additional \textquotedblleft special\textquotedblright\ invariant
which completes the basis. Based on these results, we present the following
conjecture \bigskip

\textbf{Conjecture:} \textit{(i) In dimensions $D \neq2k-1$, any curvature
invariant of order $k$ \footnote{By a curvature invariant of order $k$, we
mean a scalar constructed out of $k$ curvature tensors without any derivatives
acting on them.}, which gives second (or less) order traced field equations
can be expressed as a linear combination of the $2k$-dimensional Euler
density, the Weyl invariants and a total derivative. }

\textit{(ii) In dimensions $D=2k-1$, any curvature invariant of order $k$,
which gives second order traced field equations can be expressed as a linear
combination of the Weyl invariants \footnote{Note that the number of linearly
independent Weyl invariants of order $k$ in dimensions $D=2k-1$ is one less
than that in dimensions $D\geq2k$, due to the identity $C_{[a_{1}b_{1}%
}^{\ \ \ a_{1}b_{1}}\ldots C_{a_{k}b_{k}]}^{\ \ \ a_{k}b_{k}}=0$.}, a total
derivative and a \textquotedblleft special\textquotedblright\ invariant which
can be obtained by evaluating}
\begin{equation}
\mathcal{N}_{2k}:={\frac{1}{2^{k}}}\left(  \frac{D-2}{D-2k+1}\right)
\delta_{c_{1}d_{1}\cdots c_{k}d_{k}}^{a_{1}b_{1}\cdots a_{k}b_{k}}\left(
C^{\ \ \ c_{1}d_{1}}_{a_{1}b_{1}}\cdots C^{\ \ \ c_{k}d_{k}}_{a_{k}b_{k}%
}-R^{\ \ \ c_{1}d_{1}}_{a_{1}b_{1}}\cdots R^{\ \ \ c_{k}d_{k}}_{a_{k}b_{k}%
}\right)
\end{equation}
\textit{in $D=2k-1$.}

We now show that $\mathcal{N}_{2k}$ evaluated in $D=2k-1$ indeed gives second
order traced field equations. First, consider the following invariant of order
$k$
\begin{equation}
{\frac{1}{2^{k}}}\delta_{c_{1}d_{1}\cdots c_{k}d_{k}}^{a_{1}b_{1}\cdots
a_{k}b_{k}}\left(  C^{\ \ \ c_{1}d_{1}}_{a_{1}b_{1}}\cdots C^{\ \ \ c_{k}%
d_{k}}_{a_{k}b_{k}}-R^{\ \ \ c_{1}d_{1}}_{a_{1}b_{1}}\cdots R^{\ \ \ c_{k}%
d_{k}}_{a_{k}b_{k}}\right)  \label{termone}%
\end{equation}
Obviously, the above invariant vanishes in dimensions lower than $2k$.
However, if one expands the Weyl tensor in terms of the Riemann tensor, then
it can be factorized by $(D-2k+1)$. This can be seen as follows. Consider the
basis set of $k$-th order Riemann invariants in arbitrary dimensions. In
$D=2k-1$, not all elements of this set are linearly independent. In fact, the
basis set contains one less invariant than in $D\geq2k$. This is because of
the vanishing of the $k$-th order Lovelock density. Now, after the expanding
in terms of the Riemann tensors, the term (\ref{termone}) will not contain any
$(Riemann)^{k}$. So, this invariant cannot vanish identically in $D=2k-1$
unless it is factorized by $(D-2k+1)$.\footnote{This argument cannot be
extended to dimensions $2k-2$ since one obtains another identity involving the
Riemann invariants which is obtained be contracting the Ricci tensor with the
$(k-1)$-th order Lovelock equation.} Further expanding all the Weyl tensors,
one can convince one self that the dimensional dependence of the coefficient
of the term with $k-1$ Riemann tensors and one Ricci tensor must be
$(D-2k+1)/(D-2)$. We can now divide this factor out to get a non-vanishing
invariant in $D=2k-1$. Thus, we write the $k$th order generalization of
$\mathcal{N}_{4}$ by evaluating
\begin{equation}
{\frac{1}{2^{k}}}\left(  \frac{D-2}{D-2k+1}\right)  \delta_{c_{1}d_{1}\cdots
c_{k}d_{k}}^{a_{1}b_{1}\cdots a_{k}b_{k}}\left(  C^{\ \ \ c_{1}d_{1}}
_{a_{1}b_{1}}\cdots C^{\ \ \ c_{k}d_{k}}_{a_{k}b_{k}}-R^{\ \ \ c_{1}d_{1}%
}_{a_{1}b_{1}}\cdots R^{\ \ \ c_{k}d_{k}}_{a_{k}b_{k}}\right)
\end{equation}
in $D=2k-1$. Note that, by construction, the trace of the field equation
arising from the above invariant is of second order in all dimensions.

\section{Exact Solutions}

\label{Exact Solutions} In this section, we present exact, static solutions
for the theories defined previously. For simplicity we will first focus on the
theories having fourth order field equations, defined by an arbitrary linear
combination of the invariants $W_{1}$ and $W_{2}\,$, defined respectively in
equations (\ref{W1}) and (\ref{W2}). The theory defined by the combination
$\mathcal{N}_{6}$ in the Table I, has further interesting properties in five
dimensions, which we discuss in detail in reference \cite{OR1}. Finally we
comment on the new three dimensional theory shown in Table I, which possesses
third order traced field equations.

The class of metrics considered is:%
\begin{equation}
ds_{D}^{2}=-F\left(  R\right)  dt^{2}+\frac{dR^{2}}{G\left(  R\right)  }%
+R^{2}d\Sigma_{D-2,\gamma}^{2}\ , \label{Anz2}%
\end{equation}
where $d\Sigma_{D-2,\gamma}$ is the line element of a $\left(  D-2\right)
$-dimensional compact, orientable Euclidean manifold of constant curvature
$\gamma$. For $\gamma=1$ the manifold $\Sigma_{D-2}\ $is locally equivalent to
the sphere $S^{D-2}$, while for $\gamma=0$ it reduces to a locally flat
manifold. Finally for $\gamma=-1$ the geometry of $\Sigma_{D-2}$ is given by
the quotient $H_{D-2}/\Gamma$, where $\Gamma$ is a freely acting, discrete
subgroup of $O\left(  D-2,1\right)  $.

After a coordinate transformation and a redefinition of the arbitrary
functions, the line element (\ref{Anz2}) takes the form%
\begin{equation}
ds_{D}^{2}=N\left(  r\right)  \left[  -f\left(  r\right)  dt^{2}+\frac{dr^{2}%
}{f\left(  r\right)  }+r^{2}d\Sigma_{D-2,\gamma}^{2}\right]  \ .
\label{AnzReal}%
\end{equation}
As shown below, this gauge choice is much more convenient for our purposes.

\subsection{$C^{3}$ theories}

Here we will consider Lagrangians of the form%
\begin{equation}
\mathcal{L}=\alpha W_{1}+\beta W_{2}\ . \label{LagW1andW2}%
\end{equation}
It has been proved in \cite{DeserCn} that, for the metric (\ref{AnzReal}), the
two invariants $W_{1}$ and $W_{2}$ defined in (\ref{W1}) and (\ref{W2})
respectively, are proportional. Consequently, for a particular choice of
$\alpha/\beta$, both the Lagrangian $\mathcal{L}$ and the field equations
vanish identically. In such a situation any metric within the family
(\ref{AnzReal}) is a solution of the system. Hereafter we assume
that\ $\alpha$ and $\beta$ are generic.

\bigskip

Since, in six dimensions the gravity theories defined by combinations of
$W_{1}$ and $W_{2}$ are invariant under Weyl rescalings, let us concentrate on
this case first.

\bigskip

$%
{{}^o}%
$\textbf{ }$D=6:$ Using Weyl rescalings, one can gauge away the function
$N\left(  r\right)  $ in (\ref{AnzReal}). Then, one finds that the solution
reduces to%
\begin{equation}
ds_{6}^{2}=-\left(  ar^{2}+br+K-\frac{c\left(  1+er\right)  ^{\frac{5}{2}}%
}{r^{1/2}}\right)  dt^{2}+\frac{dr^{2}}{ar^{2}+br+K-\frac{c\left(
1+er\right)  ^{\frac{5}{2}}}{r^{1/2}}}+r^{2}d\Sigma_{4,\gamma}^{2}\ ,
\label{sol6}%
\end{equation}
where $a,b,K,c$ and $e$ are constants, which are related by%
\begin{equation}
\left\{
\begin{array}
[c]{cc}%
K=\gamma\text{ and} & c\left(  b-2Ke\right)  =0\\
\text{or} & \\
K=-\frac{1}{2}\gamma\text{ and} & c=0\text{ }%
\end{array}
\right.  \ .
\end{equation}

The Ricci scalar of this geometry diverges at $r=r_{s1}:=0$, while at
$r=r_{s2}:=-e^{-1}$ the differential scalar $\nabla_{\mu}R\nabla^{\mu}R$
diverges. For negative $e$, the region $r>r_{s2}$ must be removed from the
spacetime, since otherwise the metric is imaginary, unless $c$ vanishes. For
vanishing $c$ and $\ K=\gamma$, we obtain a conformally flat solution which
may possess one or two horizons surrounding the singularity at the origin. For
$K=-\gamma/2$, the spacetime is not conformally flat and may also describe a
black hole.

Let us note that, since in six dimensions the theory is conformally invariant,
any metric conformally related to (\ref{sol6}), will be a solution of the system.

\bigskip

$%
{{}^o}%
$\textbf{ }$D\neq6:$ For dimensions other than six, the situation is
different. Since the theory defined by (\ref{LagW1andW2}) is not invariant
under local Weyl rescaling, one naively expects the factor $N\left(  r\right)
$ in (\ref{AnzReal}) to be fixed by the field equations. However, this is not
the case, and for arbitrary dimensions, the most general solution within the
family (\ref{AnzReal}), for the theory (\ref{LagW1andW2}) is%
\begin{equation}
ds_{D}^{2}=N\left(  r\right)  \left[  -\left(  ar^{2}+br+\gamma\right)
dt^{2}+\frac{dr^{2}}{ar^{2}+br+\gamma}+r^{2}d\Sigma_{D-2,\gamma}^{2}\right]
\ , \label{ConfTempoBH}%
\end{equation}
$N(r)$ being an arbitrary function.

This can be easily seen as follows: Since in dimensions other than six, the
trace of the field equations for the theory (\ref{LagW1andW2}) is proportional
to the Lagrangian, the invariants $W_{1}\sim W_{2}$ evaluated on a solution
should vanish. For the spacetime under consideration (\ref{AnzReal}), it has
been shown in \cite{DeserCn}, that all the components of the Weyl tensor are
proportional to a single function $X$, such that the vanishing of $X$ implies
that the metric should be conformally flat. Since the restriction $W_{1}%
=W_{2}=0$, transforms covariantly under Weyl rescalings, it does not involve
the function $N\left(  r\right)  $, and the mentioned restriction reduces to
$X^{3}=0$, which implies $f\left(  r\right)  =ar^{2}+br+\gamma$. Then one is
left with a conformally flat space, and since the field equations explicitly
contain a Weyl tensor, they are fulfilled for any arbitrary function $N\left(
r\right)  $\footnote{Note that the same argument is valid for any theory with
a Lagrangian of the form $\overset{n}{\overbrace{C...C}}$ provided $D\neq2n$.
In four dimensions, for $n=3$, this solution was found in \cite{DeserC3},
where it was mentioned that the corresponding model is the simplest one that
does not admit Schwazschild horizons .}. For a smooth conformal factor
$N\left(  r\right)  $, the spacetime (\ref{ConfTempoBH}) is conformally flat
and it has been studied within the context of conformal gravity in four
dimensions in \cite{Riegert} for $\gamma=1$ and in reference \cite{Klemm} for
arbitrary $\gamma$. The three dimensional cousin of this metric, in which
$d\Sigma_{\gamma}$ is replaced by a single compact direction $d\phi$ and
$\gamma$ is an integration constant, can be obtained through ``conformal
gluing" of BTZ black holes, and is a solution of three dimensional conformal
gravity \cite{JPTempo}. In \cite{OTT} this metric was obtained within the
context of BHT new massive gravity \cite{bht, BHT2} at the special point where
the two possible maximally symmetric solutions of the theory coincide. In that
case, $\gamma$ is an arbitrary constant, the parameter $b$ plays the role of a
gravitational hair, while $a$ is fixed in terms of the coupling constant.

The metric (\ref{ConfTempoBH}) may possess an event and a Cauchy horizon,
depending on the zeros of $g_{tt}$. It generically possesses a curvature
singularity located at $r=0$, and depending on the sign of the integration
constant $a$, it represents an asymptotically locally (A)dS or flat spacetime
for ($a>0$) $a<0$ or $a=0$ respectively. The details of the different causal
structures are given in \cite{OTT}.

\subsection{The five-dimensional combination $\mathcal{N}_{6}$}

As stated in Table I, in five dimensions, there are two linearly independent
invariants whose traced field equations are of second order. Now, consider the
following linear combination as the Lagrangian
\begin{equation}
\mathcal{L}={\frac{7}{4}}W_{1}-{\frac{1}{3}}\mathcal{N}_{6}%
\end{equation}
evaluated on $D=5$. This is the unique cubic invariant for which all the
components of field equation, for static spherically symmetric spacetimes, are
of second order.

As shown in reference \cite{OR1}, the most general, non-degenerate spherically
symmetric solution, is given by
\begin{equation}
ds^{2}=-\left(  cr^{2/3}+\gamma\right)  dt^{2}+\frac{dr^{2}}{cr^{2/3}+\gamma
}+r^{2}d\Sigma_{\gamma}^{2}\ , \label{solS}%
\end{equation}
where $c$ is an integration constant and $\gamma=\pm1,0$ is the curvature of
$\Sigma_{3}$. Let us note that this spacetime is not conformally flat (it
possesses a nonvanishing Weyl tensor) unless $c=0$. For positive $c$ and
$\gamma=-1$, the metric (\ref{solS}) represents a black hole possessing an
event horizon located at located at $r_{+}=c^{-3/2}$. In this case the
geometry of the horizon is given by $H_{3}/\Gamma$, where $\Gamma$ is a freely
acting discrete subgroup$\ $of $O\left(  3,1\right)  $. The horizon hides the
curvature singularity located at $r=0$, and the asymptotic region
($r\rightarrow\infty$) is locally flat. Further interesting features of this
solution are discussed in \cite{OR1}. It is also interesting to note that
among the class of theories considered here, this is the only one which does
not admit an (A)dS solution, in the same way as the pure $K$ combination of
BHT new massive gravity \cite{bht}.

\subsection{The three dimensional case}

As shown in Table I, within the family considered, the only nontrivial theory
having third order traced field equations in three dimensions can be written
as $C_{abc}C^{abc}$, where $C_{abc}$ is the Cotton tensor. In this theory, the
most general static, spherically symmetric solution is given by%
\begin{equation}
ds^{2}=N\left(  r\right)  \left[  -\left(  ar^{2}+br-\mu\right)  dt^{2}%
+\frac{dr^{2}}{ar^{2}+br-\mu}+r^{2}d\phi^{2}\right]  \ ,
\end{equation}
where $a,b$ and $\mu$ are integration constants and $N\left(  r\right)  $ is
an arbitrary function. For smooth $N\left(  r\right)  $, as mentioned above,
this metric has an event and a Cauchy horizon, depending on the value of the parameters.

It will be interesting to study the thermodynamical properties of the black
hole within the context of AdS/CFT.

\section{Summary}

In this work, we have classified all the six-derivative Lagrangians of gravity
for which the trace of the field equations have a reduced order. We have seen
that, in dimensions greater or equal to six, when the trace of the field
equations from a generic Lagrangian is restricted to order two, we obtain an
arbitrary linear combination of three linearly independent curvature
invariants, namely, the six-dimensional Euler density and the two independent
Weyl invariants. These invariants are no longer independent in lower
dimensions due to the Schouten identities. However, in five dimensions, there
is a special invariant ${\mathcal{N}}_{6}$, which also gives field equations
with second order trace. These invariants can be used to construct interesting
cubic theories of gravity that can serve as toy models for higher derivative
theories. We have also provided a conjecture regarding all the possible
invariants of arbitrary order which gives second order traced field equations
in any dimensions. In addition, we have obtained the general spherically
symmetric solutions for a sub-class of such theories in arbitrary dimensions.
Our analysis shows that this is possible due to the reduced order of the trace
of the field equations. When the order of the trace is restricted to three, we
obtain two further invariants $\Sigma$ and $\Theta$ in arbitrary dimensions.
These two invariants are not globally independent in three and six dimensions.
In six dimensions, they reduce to the third type-B anomaly \footnote{In fact,
the anomalies are called global conformal invariants. It was first conjectured
by Deser and Schwimmer\cite{DeserSchw}, that any global conformal invariant can be expressed as
a linear combination of the Euler density and the local conformal invariants.
Recently, the conjecture has been proved by differential geometric techniques
by Alexakis \cite{Alexakis} and cohomological techniques by Boulanger \cite{Boulangercoh}. For six-derivative
invariants, in arbitrary dimensions, in addition to the two independent Weyl
invariants which are purely algebraic, there is a third local conformal
invariant which involves derivative of the curvature. Our analysis shows that
this invariant does not give field equations with reduced order trace in
arbitrary dimensions. However, they coincide (equivalent upto a total
derivative \cite{DeserSchw}) with our $\Sigma$ and $\Theta$ in dimensions
three and six.}, whereas in three dimensions they are equivalent to the square
of the cotton tensor $\sim C_{abc}C^{abc}$. We have further obtained a general
spherically symmetric solution of this theory in three dimensions.

\bigskip

\textit{Acknowledgments.} We thank Nicolas Boulanger, Hideki Maeda, David
Tempo and Steven Willison, for useful comments. This research is partially
funded by Fondecyt grants number 3095018, 11090281, and by the Conicyt grant
\textquotedblleft Southern Theoretical Physics Laboratory\textquotedblright%
\ ACT-91. The Centro de Estudios Cient\'{\i}ficos (CECS) is funded by the
Chilean Government through the Millennium Science Initiative and the Centers
of Excellence Base Financing Program of Conicyt. CECS is also supported by a
group of private companies which at present includes Antofagasta Minerals,
Arauco, Empresas CMPC, Indura, Naviera Ultragas and Telef\'{o}nica del Sur.
CIN is funded by Conicyt and the Gobierno Regional de Los R\'{\i}os.

\appendix

\section{Trace of the field equation}

\label{tofe}

Here, we prove a general property of the trace of the field equations for any
Lagrangian of the form ${\mathcal{L}}(g_{ab},R_{abcd},\nabla_{e})$ with a
fixed degree of differentiation $n$. It has been shown in \cite{Iyer:1994ys}
that the Lagrangian can always be reexpressed as
\begin{equation}
\mathcal{L}[g_{ab},R_{abcd},\nabla_{a_{1}}R_{bcde},\cdots,\nabla
_{(a_{1},\cdots,a_{p})}R_{bcde}]
\end{equation}
The field equations obtained by variation of the action with respect to the
metric takes the form
\begin{align}
&  -T^{ab}=\frac{\partial\mathcal{L}}{\partial g_{ab}}+E_{\ cde}^{a}%
R^{bcde}+2\nabla_{c}\nabla_{d}E^{acdb}+\frac{1}{2}g^{ab}\mathcal{L}\\
&  E^{bcde}=\frac{\partial\mathcal{L}}{\partial R_{bcde}}-\nabla_{a_{1}}%
\frac{\partial\mathcal{L}}{\partial\nabla_{a_{1}}R_{bcde}}+\cdots
+(-1)^{p}\nabla_{(a_{1}}\cdots\nabla_{a_{p})}\frac{\partial\mathcal{L}%
}{\partial\nabla_{(a_{1}}\cdots\nabla_{a_{p})}R_{bcde}},
\end{align}
where $T^{ab}$ is the energy-momentum tensor of the matter fields. Taking the
trace of the field equations we obtain,
\begin{equation}
-T_{\ a}^{a}=g_{ab}\frac{\partial\mathcal{L}}{\partial g_{ab}}+E^{abcd}%
R_{abcd}+\frac{D}{2}\mathcal{L}+tot.\ deriv.
\end{equation}
Now, if the Lagrangian is of fixed $n$, then it can be expressed as a linear
combination of terms of the form
\begin{equation}
\lbrack g^{..}]^{q_{1}}[R_{....}]^{q_{2}}[\nabla_{.}R_{....}]^{q_{3}}%
\cdots\lbrack\underset{p\ times}{\underbrace{\nabla_{.}\cdots\nabla_{.}}%
}R_{....}]^{q_{p+2}},
\end{equation}
such that
\begin{equation}
2q_{2}+3q_{3}+\cdots+(p+2)q_{p+2}=n.
\end{equation}

Then, under the scaling $g^{ab}\rightarrow t^{-1}g^{ab}$, $R_{bcde}\rightarrow
tR_{bcde},....,\ \overset{p\ times}{\overbrace{\nabla_{a_{1}}\cdots
\nabla_{a_{p}}}}R_{bcde}\rightarrow t\overset{p\ times}{\overbrace
{\nabla_{a_{1}}\cdots\nabla_{a_{p}}}}R_{bcde}$, the Lagrangian scales as
$\mathcal{L}\rightarrow t^{-q_{1}+q_{2}+\cdots q_{p+2}}\mathcal{L}$. However,
$q_{1}$ can be expressed in terms of other $q_{p}$'s as
\begin{align}
q_{1}  &  =\frac{1}{2}[4q_{2}+5q_{3}+\cdots+(p+4)q_{p+2}]\nonumber\\
&  =\frac{1}{2}[2(q_{2}+q_{3}+\cdots+q_{p+2})+(2q_{2}+3q_{3}+\cdots
+(p+2)q_{p+2})]\nonumber\\
&  =(q_{2}+q_{3}+\cdots+q_{p+2})+\frac{n}{2}.
\end{align}
This implies that the Lagrangian scales as $t^{-\frac{n}{2}}\mathcal{L}$. Now,
one can apply Euler's theorem of homogenous functions to write the following
relation
\begin{align}
-\frac{n}{2}\mathcal{L}  &  =-g^{ab}\frac{\partial\mathcal{L}}{\partial
g^{ab}}+R_{bcde}\frac{\partial\mathcal{L}}{\partial R_{bcde}}+\nabla_{a_{1}%
}R_{bcde}\frac{\partial\mathcal{L}}{\partial\nabla_{a_{1}}R_{bcde}}%
+\cdots+\nabla_{(a_{1}}\cdots\nabla_{a_{p})}R_{bcde}\frac{\partial\mathcal{L}%
}{\partial\nabla_{(a_{1}}\cdots\nabla_{a_{p})}R_{bcde}}\nonumber\\
&  =g_{ab}\frac{\partial\mathcal{L}}{\partial g_{ab}}+R_{bcde}\frac
{\partial\mathcal{L}}{\partial R_{bcde}}+\nabla_{a_{1}}R_{bcde}\frac
{\partial\mathcal{L}}{\partial\nabla_{a_{1}}R_{bcde}}+\cdots+\nabla_{(a_{1}%
}\cdots\nabla_{a_{p})}R_{bcde}\frac{\partial\mathcal{L}}{\partial
\nabla_{(a_{1}}\cdots\nabla_{a_{p})}R_{bcde}}\nonumber\\
&  =g_{ab}\frac{\partial\mathcal{L}}{\partial g_{ab}}+E^{bcde}R_{bcde}%
+tot.\ deriv.
\end{align}
Therefore, the trace of the field equations can be written in the form
\begin{equation}
T_{\ i}^{i}=\frac{n-D}{2}\mathcal{L}+tot.\ deriv. \label{trace}%
\end{equation}

Now, suppose that the trace of the field equations, from a Lagrangian of
$n=6$, is of third order. Then it must be some linear combination of the
invariants $L_{1},\cdots,L_{12}$. According to (\ref{trace}), in dimensions
$D\neq n$, the Lagrangian must be proportional to this combination up to a
total derivative. In dimensions $D=n$, since the trace of the field equations
is a total derivative, the only way the trace can be of at most third order is
when it identically vanishes, which is the case for conformally invariant theories.

\section{Equations of motion}

\label{eom} In this appendix, we provide the details of the analysis for the
classification presented in Table I. The equations of motion for each term in
the general Lagrangian are listed below \cite{folacci}:
\begin{align}
G_{1ab}^{(3)} &  =3R_{aq}^{\ \ ef}R_{b}^{\ qcd}R_{cdef}+6\nabla_{p}\nabla
_{q}(R_{a}^{\ qcd}R_{b\ cd}^{\ p})-\dfrac{1}{2}g_{ab}L_{1}\label{Gab1}\\
G_{2ab}^{(3)} &  =3R_{ahd}^{\ \ \ g}R_{b}^{\ prd}R_{pgr}^{\ \ \ h}-3\nabla
_{p}\nabla_{q}(R_{\ g\ h}^{p\ q}R_{a\ b}^{\ g\ h}-R_{\ hbg}^{p}R_{a}%
^{\ gqh})-\dfrac{1}{2}g_{ab}L_{2}\\
G_{3ab}^{(3)} &  =R_{acbd}R^{cspq}R_{pqs}^{\ \ \ d}-R_{a}^{\ qcd}%
R_{cdb}^{\ \ \ h}R_{qh}+R_{b}^{\ dqc}R_{adc}^{\ \ \ h}R_{qh}-\nabla_{p}%
\nabla_{q}(R_{ah}R_{b}^{\ qhp}+R_{bh}R_{a}^{\ qhp}\nonumber\\
&  +R_{\ h}^{q}R_{a\ b}^{\ h\ p}+R_{\ h}^{p}R_{a\ b}^{\ q\ h}+\dfrac{1}%
{2}(g^{pq}R_{a}^{\ hcd}R_{bhcd}+g_{ab}R^{prcd}R_{\ rcd}^{q}-g_{a}^{\ p}%
R_{b}^{\ rcd}R_{\ rcd}^{q}-g_{b}^{\ p}R_{a}^{\ rcd}R_{\ rcd}^{q}))\nonumber\\
&  -\dfrac{1}{2}g_{ab}L_{3}\\
G_{4ab}^{(3)} &  =2R_{apcd}R_{b}^{\ pcd}R+R_{ab}R^{pqcd}R_{pqcd}+\nabla
_{p}\nabla_{q}(4RR_{a\ b}^{\ q\ p}-g_{a}^{\ p}g_{b}^{\ q}R^{rscd}%
R_{rscd}+g^{pq}g_{ab}R^{rscd}R_{rscd})\nonumber\\
&  -\dfrac{1}{2}g_{ab}L_{4}\\
G_{5ab}^{(3)} &  =R_{ac}R_{b}^{\ fcd}R_{fd}+2R_{acbd}R^{cfdg}R_{fg}+\nabla
_{p}\nabla_{q}(R_{ab}R^{pq}-R_{a}^{\ p}R_{b}^{\ q}+g^{pq}R_{acbd}%
R^{cd}\nonumber\\
&  +g_{ab}R^{pcqd}R_{cd}-g_{a}^{\ p}R_{\ cbd}^{q}R^{cd}-g_{b}^{\ p}%
R_{\ cad}^{q}R^{cd})-\dfrac{1}{2}g_{ab}L_{5}\\
G_{6ab}^{(3)} &  =3R_{acbd}R^{ec}R_{e}^{\ d}+\dfrac{3}{2}\nabla_{p}\nabla
_{q}(g^{pq}R_{a}^{\ c}R_{bc}+g_{ab}R^{ep}R_{e}^{\ q}-g_{b}^{\ p}R^{qc}%
R_{ac}-g_{a}^{\ p}R^{qc}R_{bc})-\dfrac{1}{2}g_{ab}L_{6}\\
G_{7ab}^{(3)} &  =R_{ab}R^{cd}R_{cd}+2RR^{cd}R_{acbd}+\nabla_{p}\nabla
_{q}(g_{ab}g^{pq}R^{cd}R_{cd}+g^{pq}RR_{ab}-g_{a}^{\ p}g_{b}^{\ q}R^{cd}%
R_{cd}+g_{ab}RR^{pq}\nonumber\\
&  -g_{b}^{\ p}RR_{a}^{\ q}-g_{a}^{\ p}RR_{b}^{\ q})-\dfrac{1}{2}g_{ab}L_{7}\\
G_{8ab}^{(3)} &  =3R^{2}R_{ab}+3\nabla_{p}\nabla_{q}(g_{ab}g^{pq}R^{2}%
-g_{a}^{\ p}g_{b}^{\ q}R^{2})-\dfrac{1}{2}g_{ab}L_{8}\\
G_{9ab}^{(3)} &  =\nabla_{a}R\nabla_{b}R-2\Box RR_{ab}-2(g_{ab}g_{cd}%
-g_{ac}g_{bd})\nabla^{c}\nabla^{d}\Box R-\dfrac{1}{2}g_{ab}L_{9}\\
G_{10ab}^{(3)} &  =\nabla_{a}R^{cd}\nabla_{b}R_{cd}+2\nabla_{c}R_{a}%
^{\ d}\nabla^{c}R_{bd}-\Box^{2}R_{ab}-\nabla_{c}\nabla_{d}\Box R^{cd}%
g_{ab}+\nabla_{a}\nabla_{c}\Box R_{\ b}^{c}+\nabla_{b}\nabla_{c}\Box
R_{\ a}^{c}\nonumber\\
&  -2R_{acbd}\Box R^{cd}+2R_{c(a}\Box R_{\ b)}^{c}+2\nabla_{c}[R_{d}%
^{\ c}\nabla_{(b}R_{a)}^{\ d}-R_{d(a}\nabla^{c}R_{\ b)}^{d}-R_{d(b}\nabla
_{a)}R^{cd}]-\dfrac{1}{2}g_{ab}L_{10}\\
G_{11ab}^{(3)} &  =2G_{1ab}^{(3)}-4G_{2ab}^{(3)}+2G_{3ab}^{(3)}-4G_{5ab}%
^{(3)}+4G_{6ab}^{(3)}-G_{9ab}^{(3)}+4G_{10ab}^{(3)}\\
G_{12ab}^{(3)} &  =G_{5ab}^{(3)}-G_{6ab}^{(3)}+\frac{1}{4}G_{9ab}%
^{(3)}\ .\label{Gab8}%
\end{align}
Therefore the trace of the full field equations can be expressed as
\begin{align}
A^{i}G_{ia}^{(3)a} &  =(3-D/2)A^{i}L_{i}+\nabla_{p}\nabla_{q}[(6A^{1}%
+3A^{2}-{\frac{D-2}{2}}A^{3}-4A^{11})R^{pabc}R_{\ abc}^{q}\nonumber\\
&  +(-3A^{2}-2A^{3}+(D-2)A^{5}+2(D-2)A^{10}+24A^{11}+(D-2)A^{12}%
)R_{ab}R^{apbq}\nonumber\\
&  +(-2A^{3}-A^{5}+{\frac{3(D-2)}{2}}A^{6}-2A^{10}+16A^{11}-(D+1)A^{12}%
)R_{\ a}^{p}R^{qa}\nonumber\\
&  +(4A^{4}+A^{5}+(D-2)A^{7}-(D-4)A^{10}-12A^{11}-{\frac{D-8}{2}}%
A^{12})RR^{pq}\nonumber\\
&  +(-{\frac{1}{2}}A^{3}+(D-1)A^{4}-A^{11})g^{pq}R^{abcd}R_{abcd}\nonumber\\
&  +(A^{5}+{\frac{3}{2}}A^{6}+(D-1)A^{7}-{\frac{D}{2}}A^{10}-10A^{11}%
-{\frac{1}{2}}A^{12})g^{pq}R^{ab}R_{ab}\nonumber\\
&  +(A^{7}+3(D-1)A^{8}-A^{9}+{\frac{D-4}{4}}A^{10}+3A^{11}+{\frac{D-8}{8}%
}A^{12})g^{pq}R^{2}\nonumber\\
&  +(-2(D-1)A^{9}-{\frac{D}{2}}A^{10}-2A^{11}-{\frac{D-1}{2}}A^{12})g^{pq}\Box
R]
\end{align}
Now we impose the trace to be proportional to the Lagrangian. This in turn
requires the second term on the right hand side to vanish. To realize this,
one has to choose the coefficients in such a way that, the symmetric tensor
quadratic in curvature inside the operator $\nabla_{p}\nabla_{q}$ is
proportional to the Gauss-Bonnet field equations. This gives us a set of $8$
equations in $12$ variables and one arbitrary parameter $u$. They are
\begin{align}
&  6A^{1}+3A^{2}-{\frac{D-2}{2}}A^{3}-4A^{11}=-2u,\nonumber\\
&  -{\frac{1}{2}}A^{3}+(D-1)A^{4}-A^{11}=u/2,\nonumber\\
&  -3A^{2}-2A^{3}+(D-2)A^{5}+2(D-2)A^{10}+24A^{11}+(D-2)A^{12}=4u,\nonumber\\
&  -2A^{3}-A^{5}+{\frac{3(D-2)}{2}}A^{6}-2A^{10}+16A^{11}-(D+1)A^{12}%
=4u,\nonumber\\
&  4A^{4}+A^{5}+(D-2)A^{7}-(D-4)A^{10}-12A^{11}-{\frac{D-8}{2}}A^{12}%
=-2u,\nonumber\\
&  A^{5}+{\frac{3}{2}}A^{6}+(D-1)A^{7}-{\frac{D}{2}}A^{10}-10A^{11}-{\frac
{1}{2}}A^{12}=-2u,\nonumber\\
&  A^{7}+3(D-1)A^{8}-A^{9}+{\frac{D-4}{4}}A^{10}+3A^{11}+{\frac{D-8}{8}}%
A^{12}=u/2,\nonumber\\
&  -2(D-1)A^{9}-{\frac{D}{2}}A^{10}-2A^{11}-{\frac{D-1}{2}}A^{12}%
=0.\label{SystemD}%
\end{align}
The matrix of linear equations has rank $8$, which implies that the general
solution can be written in terms of $5$ arbitrary parameters $x,y,z,u$ and
$v$. In $D>5$, the solution is given as%

\begin{align}
&  A^{1}={\frac{1 }{12}}\left[  2(D^{2}+5D-10)x+2(D-2)^{2}y-6(3D+2)z+(D^{2}%
-4)v+3(D-2)u\right]  ,\nonumber\\
&  A^{2}=-\frac{1}{6}\left[  8(2D-3)x+2(D-2)^{2}y-8(2D+3)z+(D^{2}%
-4)v+4(D-1)u\right] \nonumber\\
&  A^{3}=2(D-1)x-2z-u\nonumber\\
&  A^{4}=x\nonumber\\
&  A^{5}=-4x-(D-2)y+\frac{8(D-1)}{D-2}z-\frac{D}{2}v-2u\nonumber\\
&  A^{6}=\frac{1}{3}\left[  8x-2y-\frac{24}{D-2}z+v\right] \nonumber\\
&  A^{7}=y\nonumber\\
&  A^{8}=-\frac{1 }{24(D-1)^{2}}\left[  8(D-1)y+16Dz-(D^{2}%
-D+2)v-4(D-1)u\right] \nonumber\\
&  A^{9}=\frac{1}{4(D-1)(D-2)}\left[  8z+(D-2)v\right] \nonumber\\
&  A^{10}=-\frac{4}{D-2}z-v\nonumber\\
&  A^{11}=z\nonumber\\
&  A^{12}=v.
\end{align}

In dimensions $D>5$, one can apply the following transformation
\begin{align}
&  x\rightarrow\frac{3}{2(D-2)(D-1)}\left[  2(D-2)(D-1)a+4b+c\right]
,\nonumber\label{reparam}\\
&  y\rightarrow-{\frac{3}{(D-2)^{3}(D-1)}}\left[  4(D-2)^{3}%
(D-1)a+8(2D-3)b+(D^{2}+D-4)c\right]  ,\nonumber\\
&  z\rightarrow d,\nonumber\\
&  u\rightarrow6(D-5)a,\nonumber\\
&  v\rightarrow24e.
\end{align}
such that
\[
A^{i}L_{i}=a\ \mathcal{E}_{6}+b\ W_{1}+c\ W_{2}+d\ \Sigma+e\ \Theta.
\]
where
\begin{align}
\Sigma=  &  -\frac{1}{2}(3D+2)L_{1}+\frac{4}{3}(2D+3)L_{2}-2L_{3}%
+\frac{8(D-1)}{D-2}L_{5}-\frac{8}{D-2}L_{6}-\frac{2D}{3(D-1)^{2}}%
L_{8}\nonumber\\
&  +\frac{2}{(D-2)(D-1)}L_{9}-\frac{4}{D-2}L_{10}+L_{11}\nonumber\\
=  &  -\frac{1}{2}(3D-2)L_{1}+\frac{8D}{3}L_{2}+\frac{4D}{D-2}L_{5}%
+\frac{4(D-4)}{D-2}L_{6}-\frac{2D}{3(D-1)^{2}}L_{8}-\frac{D(D-3)}%
{(D-2)(D-1)}L_{9}\nonumber\\
&  +\frac{4(D-3)}{D-2}L_{10}+total\ derivative\label{Sigma}\\
\Theta=  &  2(D^{2}-4)L_{1}-4(D^{2}-4)L_{2}-12DL_{5}+8L_{6}+\frac{D^{2}%
-D+2}{(D-1)^{2}}L_{8}+\frac{6}{D-1}L_{9}-24L_{10}+24L_{12}\nonumber\\
=  &  2\left(  D^{2}-4\right)  L_{1}-4\left(  D^{2}-4\right)  L_{2}%
-12(D-2)L_{5}-16L_{6}+\frac{D^{2}-D+2}{(D-1)^{2}}L_{8}+\frac{6D}{D-1}%
L_{9}\nonumber\\
&  -24L_{10}+total\ derivative. \label{Theta}%
\end{align}
Note that the determinant of the transformation (\ref{reparam}) is
${\frac{2592(D-5)}{(D-2)^{3}(D-1)}}$. Now in $D\leq5$, one needs to solve the
system of equations (\ref{SystemD}) for each value of $D$ separately. The
Lagrangians obtained are tabulated in Table $1$.


\begin{thebibliography}{99}                                                                                               %


\bibitem {Lovelock}D.~Lovelock,
J.\ Math.\ Phys.\ \textbf{12} (1971) 498.


\bibitem {BD}D.~G.~Boulware and S.~Deser,
Phys.\ Rev.\ Lett.\ \textbf{55}, 2656 (1985).


\bibitem {JTW}J.~T.~Wheeler,
Nucl.\ Phys.\ B \textbf{273}, 732 (1986).


\bibitem {GG}C.~Garraffo and G.~Giribet,
Mod.\ Phys.\ Lett.\ A \textbf{23}, 1801 (2008) [arXiv:0805.3575 [gr-qc]].


\bibitem {CHARM}C.~Charmousis,
Lect.\ Notes Phys.\ \textbf{769}, 299 (2009) [arXiv:0805.0568 [gr-qc]].


\bibitem {WITT}E.~Witten,
Adv.\ Theor.\ Math.\ Phys.\ \textbf{2}, 505 (1998) [arXiv:hep-th/9803131].


\bibitem {ZEG}R.~Zegers,
J.\ Math.\ Phys.\ \textbf{46}, 072502 (2005) [arXiv:gr-qc/0505016].


\bibitem {Riegert}R.~J.~Riegert,
Phys.\ Rev.\ Lett.\ \textbf{53}, 315 (1984).


\bibitem {birk}J. Oliva and S. Ray. To appear

\bibitem {bht}E.~A.~Bergshoeff, O.~Hohm and P.~K.~Townsend,
Phys.\ Rev.\ Lett.\ \textbf{102}, 201301 (2009) [arXiv:0901.1766 [hep-th]].


\bibitem {OTT}J.~Oliva, D.~Tempo and R.~Troncoso,
JHEP \textbf{0907}, 011 (2009) [arXiv:0905.1545 [hep-th]].


\bibitem {OR1}J.~Oliva and S.~Ray,
arXiv:1003.4773 [gr-qc].


\bibitem {QTG}R.~C.~Myers and B.~Robinson,
JHEP \textbf{1008}, 067 (2010) [arXiv:1003.5357 [gr-qc]].


\bibitem {QTGHOL}R.~C.~Myers, M.~F.~Paulos and A.~Sinha,
JHEP \textbf{1008}, 035 (2010) [arXiv:1004.2055 [hep-th]].


\bibitem {farhoudi}M.~Farhoudi,
Gen.\ Rel.\ Grav.\ \textbf{41}, 117 (2009) [arXiv:gr-qc/9510060].


\bibitem {naoda}M.~Nakasone and I.~Oda,
Prog.\ Theor.\ Phys.\ \textbf{121}, 1389 (2009) [arXiv:0902.3531 [hep-th]].


\bibitem {BHT2}E.~A.~Bergshoeff, O.~Hohm and P.~K.~Townsend,
Phys.\ Rev.\ D \textbf{79}, 124042 (2009) [arXiv:0905.1259 [hep-th]].


\bibitem {fulling}S.~A.~Fulling, R.~C.~King, B.~G.~Wybourne and
C.~J.~Cummins,
Class.\ Quant.\ Grav.\ \textbf{9} (1992) 1151.


\bibitem {Cites On Anomaly1}

L.~Bonora, P.~Cotta-Ramusino and C.~Reina,
Phys.\ Lett.\ B \textbf{126}, 305 (1983).


\bibitem {Cites On Anomaly2}L.~Bonora, P.~Pasti and M.~Bregola,
Class.\ Quant.\ Grav.\ \textbf{3}, 635 (1986).


\bibitem {Cites On Anomaly3}S.~Deser and A.~Schwimmer,
Phys.\ Lett.\ B \textbf{309}, 279 (1993) [arXiv:hep-th/9302047].


\bibitem {DeserCn}S.~Deser and A.~V.~Ryzhov,
Class.\ Quant.\ Grav.\ \textbf{22}, 3315 (2005) [arXiv:gr-qc/0505039].


\bibitem {DeserC3}S.~Deser and B.~Tekin,
Class.\ Quant.\ Grav.\ \textbf{20}, 4877 (2003) [arXiv:gr-qc/0306114].


\bibitem {Klemm}D.~Klemm,
Class.\ Quant.\ Grav.\ \textbf{15}, 3195 (1998) [arXiv:gr-qc/9808051].


\bibitem {JPTempo}J.~Oliva, D.~Tempo and R.~Troncoso,
Int.\ J.\ Mod.\ Phys.\ A \textbf{24}, 1588 (2009) [arXiv:0905.1510 [hep-th]].


\bibitem {Iyer:1994ys}V.~Iyer and R.~M.~Wald,
Phys.\ Rev.\ D \textbf{50}, 846 (1994) [arXiv:gr-qc/9403028].


\bibitem {folacci}Y.~Decanini and A.~Folacci,
Class.\ Quant.\ Grav.\ \textbf{24}, 4777 (2007) [arXiv:0706.0691 [gr-qc]].


\bibitem {DeserSchw}S.~Deser and A.~Schwimmer,
Phys.\ Lett.\  B {\bf 309}, 279 (1993)
[arXiv:hep-th/9302047].


\bibitem {Boulangercoh}N.~Boulanger,
JHEP {\bf 0707}, 069 (2007)
[arXiv:0704.2472 [hep-th]].


\bibitem {Alexakis}Spyros Alexakis, "The decomposition of Global Conformal Invariants: On a conjecture of Deser and Schwimmer", 0711.1685v1
\end{thebibliography}
\end{document}